%% file: Paper.tex
\newcommand{\CLASSINPUTtoptextmargin}{0.75in}%
\newcommand{\CLASSINPUTbottomtextmargin}{1in}%
\newcommand{\CLASSINPUTinnersidemargin}{0.63in}%
\newcommand{\CLASSINPUToutersidemargin}{0.63in}%
\documentclass[conference,10pt,letterpaper]{IEEEtran}% 
\usepackage{amsmath}% for double integral symbol in this template
\usepackage{graphicx}% for figures
\usepackage{siunitx}
\usepackage{multirow}% to allow multiple-row elements in tabular environment
\usepackage[none]{hyphenat}% turn off hyphenation to make text extraction and indexing easier
\usepackage{float}% better control of floating figures and tables
\usepackage{subfig}% for subfigures within figures
\usepackage{dblfloatfix}% fix to allow page-wide floats at bottom of page
\usepackage{tikz}
\usepackage{textcomp}
\usepackage{hyperref}
\usepackage{lipsum}

\newcommand\copyrighttext{%
  \footnotesize \textcopyright 2012 IEEE. Personal use of this material is permitted.
  Permission from IEEE must be obtained for all other uses, in any current or future
  media, including reprinting/republishing this material for advertising or promotional
  purposes, creating new collective works, for resale or redistribution to servers or
  lists, or reuse of any copyrighted component of this work in other works.
  }
\newcommand\copyrightnotice{%
\begin{tikzpicture}[remember picture,overlay]
\node[anchor=south,yshift=10pt] at (current page.south) {\fbox{\parbox{\dimexpr\textwidth-\fboxsep-\fboxrule\relax}{\copyrighttext}}};
\end{tikzpicture}%
}

\usepackage{t1enc}% allows access to various special characters
\usepackage{times}% change font to Nimbus Roman (based on Times Roman)
\input{IMS_modify_IEEEtran_18b_CTAN_V6}

\begin{document}
%%%%%%%%%%%%%%%%%%%%%%%%%%%%%%%%%%%%%%%%%%%%%%%%%%%%%%%%%%%%%%%%%%%%%%%%%%%%%
% We use \raggedbottom to avoid latex adding vertical space around headings.
% This gives a better idea to the author about how much white space remains
% as the page limit is approached.
\raggedbottom
%
%%%%%%%%%%%%%%%%%%%%%%%%%%%%%%%%%%%%%%%%%%%%%%%%%%%%%%%%%%%%%%%%%%%%%%%%%%%%%
% PAPER TITLE AND AUTHOR BLOCK
%
% The paper title can use linebreaks \\ within to get better formatting if desired.
%
% \title{Low-Loss EBG-Bends for 3D-Printed Dielectric Image Lines\\in sub-THz Applications}
\title{3D-Printed Dielectric Image Lines towards\\ Chip-to-Chip Interconnects for subTHz-Applications}
%
% Next we define the author names and affiliations.
% Author names are listed using \IMSauthorblockNAME{} with comma separators between names.
% Affiliations are listed using \IMSauthorblockAFFIL{} with \\ separators between affiliations.
% Email addresses are listed using \IMSauthorblockEMAIL{} with comma separators between emails.
% See below for examples of each of these.
%
% Symbols marking author-affiliation relations are output using \IMSauthorrefmark{}.
%
% Next we typeset the authorblock either as visible text, or as an empty
% box of the same size, based on the value of the Blind Review Flag.
% Note that the Blind Review Flag also determines whether the Acknowledgements
% section is visible or invisible.
% To set the flag to Blind Review mode, simply uncomment the next line
% \IMSthispaperforblindreview
% or to set the flag to Final Paper mode (with author block visible) then
% simply uncomment the next line:
\IMSthispaperforfinalpublication
\IMSauthor{%
\IMSauthorblockNAME{% Author Names
Leonhard Hahn\IMSauthorrefmark{\#1},
Tim Pfahler\IMSauthorrefmark{\#},
Tobias Bader\IMSauthorrefmark{\#},
Gerald Gold\IMSauthorrefmark{\#},
Martin Vossiek\IMSauthorrefmark{\#} and 
Christian Carlowitz\IMSauthorrefmark{\#}
}% end of \IMSauthorblockNAME
\\%
\IMSauthorblockAFFIL{% Author Affiliations
\IMSauthorrefmark{\#}Institute of Microwaves and Photonics (LHFT), Friedrich-Alexander-Universität Erlangen-Nürnberg
}% end of \IMSauthorblockAFFIL
\\%
\IMSauthorblockEMAIL{% Author Emails
\IMSauthorrefmark{1}leonhard.hahn@fau.de
}% end of \IMSauthorblockEMAIL
}% end of \IMSauthor
%
% Next we make the title/author block using the information defined above.
\maketitle
%
% trigger copyright notice here
\copyrightnotice

\begin{abstract}
This paper reports on 3D-printed dielectric image lines for low-loss subTHz applications between 140~and~220\,GHz. In contrast to conventional dielectric waveguides, a conductive copper~substrate is used to achieve robust routing and increased mechanical stability. For easy integration and characterization of the dielectric image line within a waveguide measurement setup, a low-loss mode-converter for flexible mounting is further designed. The characterized overall system exhibits a broadband match of at least 20\,dB over the entire frequency band, with minimal losses of below 0.35\,dB/cm. Furthermore, multi-line characterization is performed for de-embedding the propagation parameters \boldmath{$\alpha$} and \boldmath{$\beta$} of both the dielectric transmission line and the mode-converter, and finally, the influence of discontinuities such as bending radii on the transmission behavior is evaluated. Due to the simplicity of the underlying 3D-printing technology, the proposed concept features extremely low cost and complexity, yet offers high flexibility and outperforms the losses of conventional transmission lines. 
\end{abstract}


\section{Introduction}
The mmW- and subTHz-domain presents an extremely promising region for further improvement of the performance of modern RF systems. The herefore associated possibility of substantial increase in frequency and bandwidth benefits the resolution capability of imaging radar systems or the data rate of communication systems, therefore motivating the development of a large number of subTHz-capable FMCW systems in recent years \cite{Yi.2020,Mangiavillano.2022,Statnikov.2013}. Current investigations clearly indicate the potential and the necessity of those subTHz-FMCW radar applications, especially in the area of high-resolution close-range imaging \cite{Jasteh.2015,Gu.2021}. However, the transmission loss arising at high frequencies is significant, thus preventing the use of conventional planar transmission lines such as microstrip lines (MSL) or coplanar waveguides (CPW) for signal distribution.
Other types of non-planar transmission lines, such as RF cables or hollow waveguides, either dont support the required frequency range any more or have the disadvantage of low flexibility and extremly high manufacturing efforts. Especially in the case of high-performance multi-channel imaging systems with several integrated, but spatially distributed receiver circuits (ICs), the lack of low-loss transmission options is a serious problem due to their equal dependency on mixers in the receive path, which consequently require large distribution networks for feeding every IC with one common local oscillator's signal (LO). Dielectric image lines (DIL) represent one way to realize such subTHz distribution networks. Compared to conventional planar transmission lines, these provide an extremely low-loss \cite{King.1957}, hereby enabling the realization of longer, off-chip connections even at subTHz frequencies \cite{Torabi.2021}.
\begin{figure}[t]
\centering
\includegraphics[width=90mm]{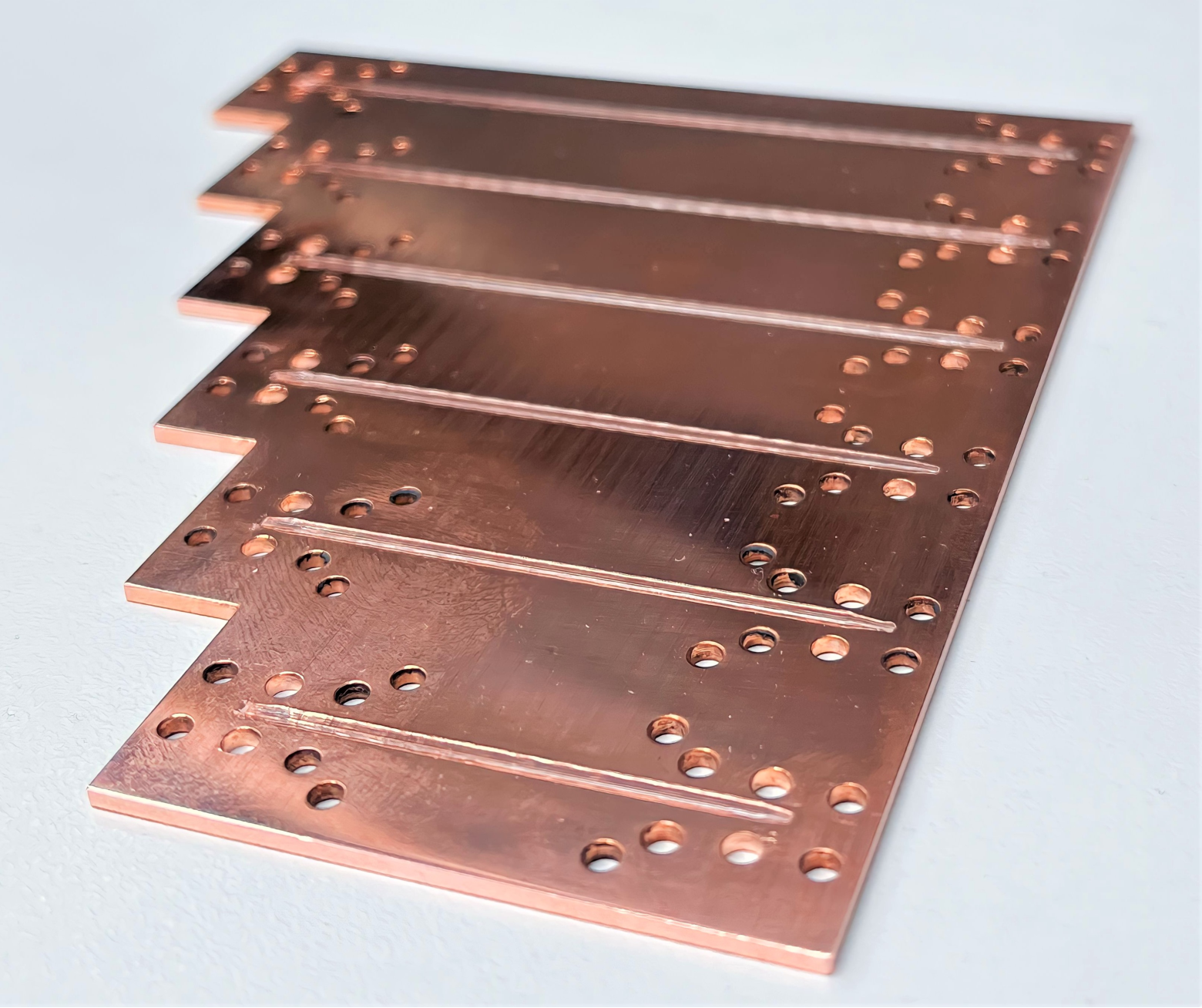}
\caption{3D-printed dielectric image lines on top of a copper substrate.}
\label{fig:DIL_Lines}
\end{figure}
In addition to large LO distribution networks, direct chip-to-chip interconnects \cite{Holloway.2017} or on-chip antennas \cite{Tesmer.2022} can be created. FurthermoreF, the possibility of additive manufacturing (3D-printing) makes DILs extremely lightweight, cost-effective and flexible to manufacture \cite{Distler.2019}, which is a significant advantage over extremely expensive RF cables or conventional metallic waveguides. For this reason, dielectric waveguides have recently become a greater focus of study and part of a wide variety of investigations \cite{Distler.2018,Fukuda.2011,Geiger.2017}. The DIL is to be understood as a special form of dielectric waveguide, since an additional conductive surface is used in addition to the dielectric \cite{King.1958}, see Fig.~\ref{fig:DIL_Lines}. This serves as an anchor for the DIL, which can be very fragile at high frequency, thus improving the guidance and mechanical stability of the system.

This paper therefore presents exclusively 3D-printed dielectric image lines for G-band applications from \SI{140}{\giga\hertz} up to \SI{220}{\giga\hertz}. Starting from the underlying theory in Section~\ref{Chapter:Theory}, the design and fabrication process of both the transmission line and its suitable mode-converter is discussed in Section~\ref{Chapter:Measurement_Setup}, enabling the connection and use of the DIL in a waveguide system. Finally, Section~\ref{Chapter:Results} contains the characterization of the resulting overall concept.

\section{Image Line Topology}\label{Chapter:Theory} 
\begin{figure}[htb]
\centering
\includegraphics[width=70mm]{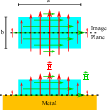}
\caption{Electric and magnetic field distribution of a conventional dielectric waveguide (top) and its corresponding image line (bottom).}
\label{fig:DIL_Fields}
\end{figure}

The basic development of dielectric image lines reaches far back \cite{King.1958,Solbach.1978} and is to be understood as a modification of the well known conventional dielectric waveguide. Due to its relative permitivitty $\varepsilon_{\text{r}} > 1$, the basic dielectric waveguide (DWG) transfers electromagnetic power with its dominant $HE_{11}$ mode axially along its dielectric, with radially decreasing electric~($\Vec{E}$) and magnetic~($\Vec{H}$) field components \cite{Distler.2019}. Note that in contrast to hollow waveguides, dielectric based waveguides have frequency dependent field components outside their transmission line. For a well-defined polarization state of the guided waves, the DWG geometries are supposed to offer an aspect ratio of $a/b = 2$ \cite{Hofmann.2003}, see Fig.~\ref{fig:DIL_Fields}. DILs are based on the exact same operating principle, but unlike conventional DWGs, they use a conductive sheet as mechanical anchor on their bottom side, enhancing their routing, their mechanical stability and reducing their physical size. In addition, the conducting sheet acts as a polarization anchor, hence simplifies the mode conversion problem by preventing the generation of parasitic modes \cite{King.1958}.
To further ensure low-loss propagation of electromagnetic waves, both the electric and magnetic boundary condition of the metal surface ($\Vec{E_{\text{tan}}}=0$, $\Vec{H_{\text{0}}}=0$) must comply with the field distribution of the dominant $HE_{11}$ mode. This is achieved by inserting the metal sheet into the DWG's~Y-symmetry~plane (image plane), allowing the E- and H-field components to remain unchanged and changing the aspect ratio to $a/b = 4$. The resulting field distribution of both the DWG and its corresponding image line can be seen in Fig.~\ref{fig:DIL_Fields}.
In this work, the DIL's dielectric is Cyclic Olefin Copolymer (COC) with a relative permittivity of $\varepsilon_{\text{r}}=2.2$. This material is very well suited due to its low dissipation~factor of $\tan \delta = 1 \cdot 10^{-4}$, its low water absorption rate and finally its simple 3D-printability with common FDM processes.

\section{Measurement Setup}\label{Chapter:Measurement_Setup} 
\begin{figure}[t]
\centering
\includegraphics[width=91mm]{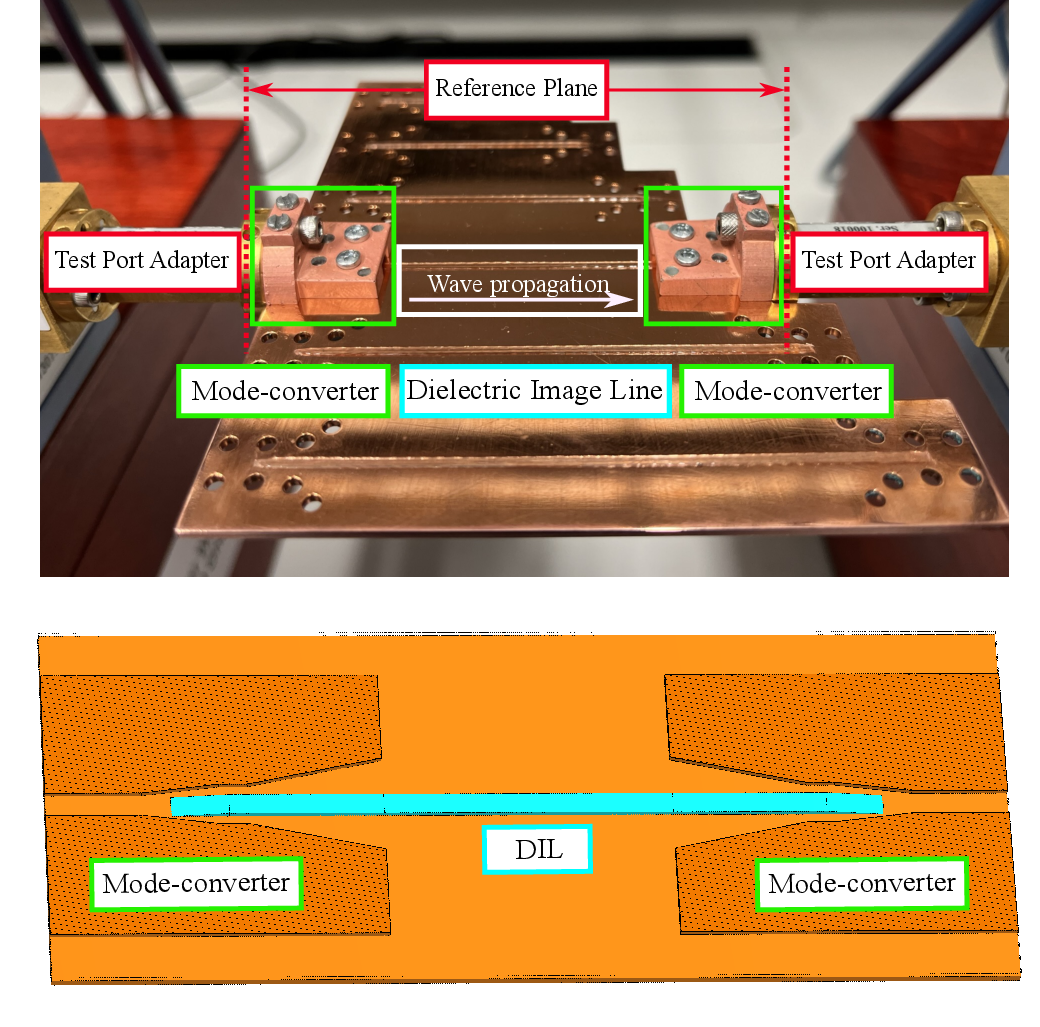}
\caption{Measurement setup for the back-to-back characterization of DIL system (top) with sectional view of its schematic (bottom).}
\label{fig:Measurement_setup}
\end{figure}
Since the dominant DIL mode $HE_{11}$ does not occur in any common type of transmission line, the design of a mode-converter is necessary. 
Key aspect of the design is the possibility to flexibly mount the mode-converter on the upper side of the image surface and thus to excite the DIL, which is lying directly underneath, for subsequent measurements. The mode-converter, like all the DILs, was initially manufactured using 3D-printing technologies, however, a CNC-milled copper version has proven better suitability in respect to mounting precision and is therefore used in this work.
\begin{figure}[b]
\centering
\includegraphics[width=85mm]{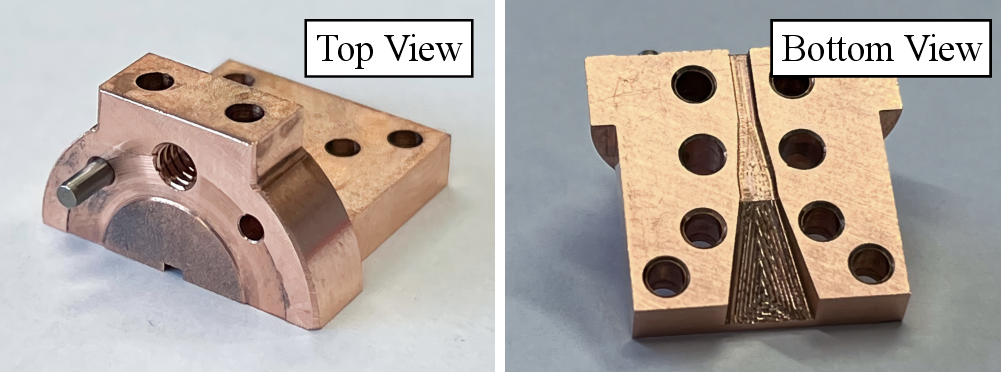}
\caption{CNC-milled mode-converter in split-block technology with WR05 waveguide flange for $HE_{11}$ mode excitation of the 3D-printed DIL.}
\label{fig:Measurement_setup_detailed}
\end{figure}
To generate the desired signals and feed the mode-converters in the G-band (\SI{140}{\giga\hertz}~to~\SI{220}{\giga\hertz}), the ZVA-Z220 frequency converters from Rohde~\&~Schwarz were used, which are fed by the ZVA24~network~analyzer. The DILs are located on the top of the mirror surface after 3D-printing and connect two mode-converters to form a symmetrical back-to-back setup, so that both reflection- and transmission measurements can be performed to characterize the DILs thoroughly. The schematic concept with mode-converters and DIL as a sectional view is illustrated in Fig.~\ref{fig:Measurement_setup}, the resulting measurement setup is shown in Fig.~\ref{fig:Measurement_setup}. After TOSM~calibration, the reference plane is located exactly at the mode-converters input, so that a measurement of the entire system is possible. Due to the strong similarity of the injected waveguide mode $TE_{10}$ to the desired hybrid DIL mode $HE_{11}$, the mode-converter only consists of a slightly modified, double-tapered waveguide. After assembly of the mode-converters onto the metal sheet, the printed DILs are located directly in the waveguide channel, see Fig.~\ref{fig:Measurement_setup_detailed}.

\begin{figure}[b]
\centering
\includegraphics[width=85mm]{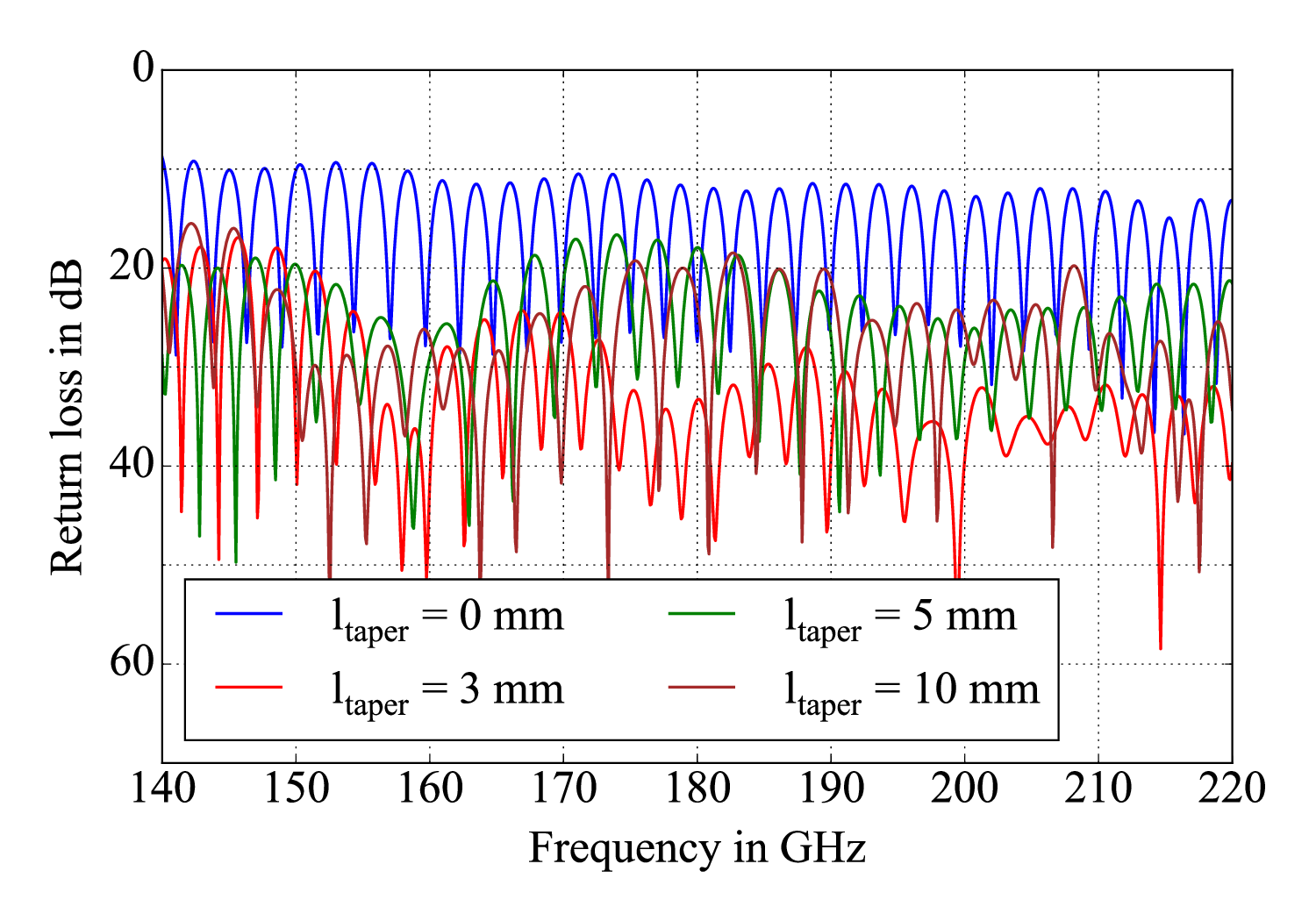}
\caption{Influence of the DIL's initial tapering length on their simulated return loss.}
\label{fig:DIL_Taper}
\end{figure}

The first taper only applies to the width of this waveguide and serves to simplify the subsequent assembly, which otherwise can become challenging due to the similar width-dimensions of DIL and mode-converter's waveguide channel. This conflict worsens with the rather high printing tolerance of the used FDM processes. The second taper includes both the width and the height dimension and thus transforms the waveguide channel into a structure comparable to a horn antenna, which contributes to matching improvements of the DIL. To further enhance matching, the DIL must be designed with a linearly increasing taper of its height. The achievable matching is heavily dependent on the taper length, as can be seen in Fig.~\ref{fig:DIL_Taper}a). Based on this results, a length of $l_{\text{taper}} = \SI{3}{\milli\meter}$ was chosen.
All geometric dimensions of the mode-converter and the DIL have been simulated and optimized for best return loss using electromagnetic field simulations in CST Microwave Studio. This simulations also reveal that compared to other transmission line types, the DIL-topology has a fairly high tolerance for almost all geometric deviations, with the only exception being its initial height tapering.

\section{Results}\label{Chapter:Results} 
\begin{figure}[t]
\centering
\includegraphics[width=85mm]{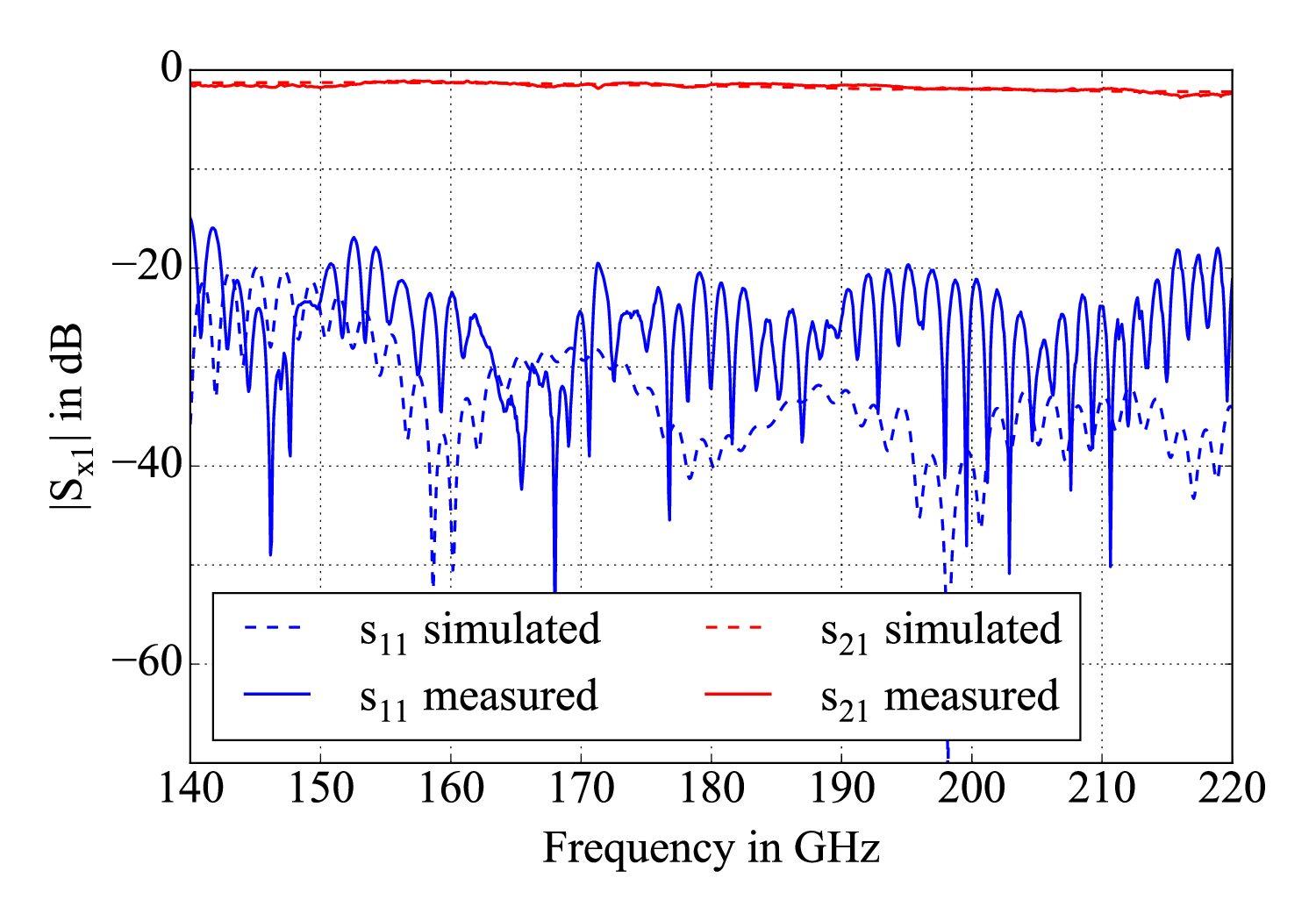}
\caption{Simulated and measured S-parameters of a single DIL with length $l = \SI{60}{\milli\meter}$.}
\label{fig:DIL_SingleLine_SimMeas}
\end{figure}
For a thorough characterization of the 3D-printed dielectric image lines, S-parameter measurements are performed after the TOSM calibration using the measurement setup presented in Section~\ref{Chapter:Measurement_Setup}. This allows for determination of the transmission- and reflection properties and further investigation of relevant characteristics such as the complex propagation coefficient $\gamma=\alpha+j\beta$. Fig.~\ref{fig:DIL_SingleLine_SimMeas} shows the simulated and measured S-parameters for a DIL of length $l=\SI{60}{\milli\meter}$, the influence of both mode-converters is included. It is clearly evident that the DIL is very well matched over the entire frequency band from \SI{140}{\giga\hertz} to \SI{220}{\giga\hertz} with a return loss (RL) of constantly approx. \SI{20}{\decibel}. Furthermore, it exhibits an outstandingly low insertion loss of below \SI{2}{\decibel} for frequencies up to \SI{200}{\giga\hertz}, while values of \SI{2.8}{\decibel} are still achieved for the maximum frequency of \SI{220}{\giga\hertz}.
\begin{figure}[b]
\centering
\includegraphics[width=85mm]{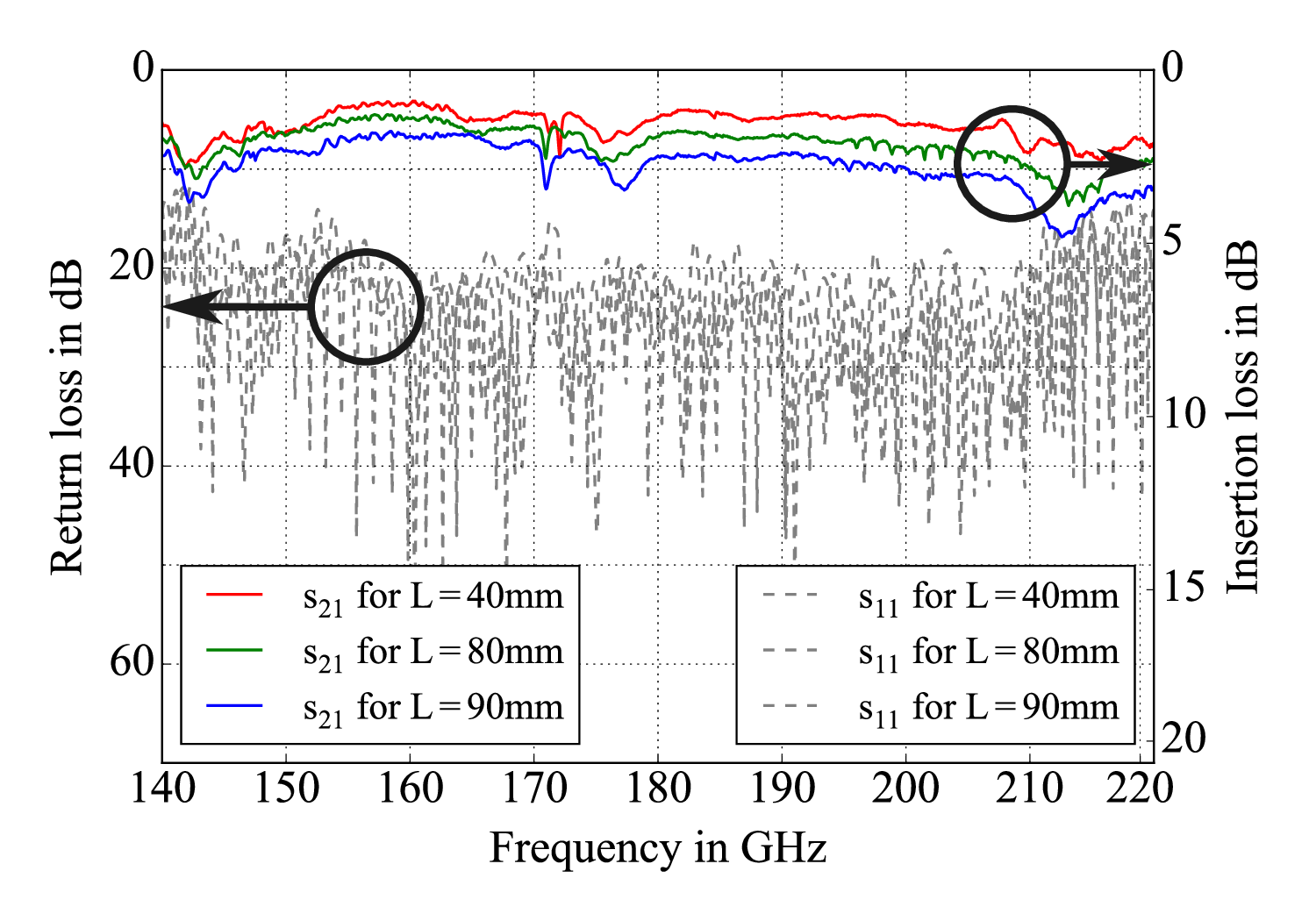}
\caption{Measured S-parameters of a multi-line DIL for estimation of the propagation coefficient $\gamma$.}
\label{fig:DIL_MultiLine}
\end{figure}
The measured IL values are in excellent agreement with the simulation data, whereas the RL values deviate slightly from the simulation. This can be explained by the imprecision of the FDM 3D-printing technology, which makes precise manufacturing of the DIL's taper very challenging. Deviations in taper length and -structure have a fairly high effect on the return loss of the DILs, as previously shown in Fig.~\ref{fig:DIL_Taper}. To de-embed the true properties of the transmission lines only, multi-line measurements are performed using DILs of different lengths, between \SI{40}{\milli\meter} and \SI{90}{\milli\meter}). With constant influence of the two mode-converters, the resulting deviations of the $s_{21}$-parameters  allow the extraction of the propagation coefficient $\gamma_{\text{DIL}}$, from which the DIL's attenuation coefficient $\alpha_{\text{DIL}}$ and phase coefficient $\beta_{\text{DIL}}$ can easily be read. Fig.~\ref{fig:DIL_MultiLine} exemplarily shows the resulting return- and insertion loss of the multi-line measurement for DILs of three different lengths. As expected, the rise of measured insertion loss is clearly evident with increasing line length, while the measured return loss remains independent of length. 
\begin{figure}[h]
\centering
\includegraphics[width=85mm]{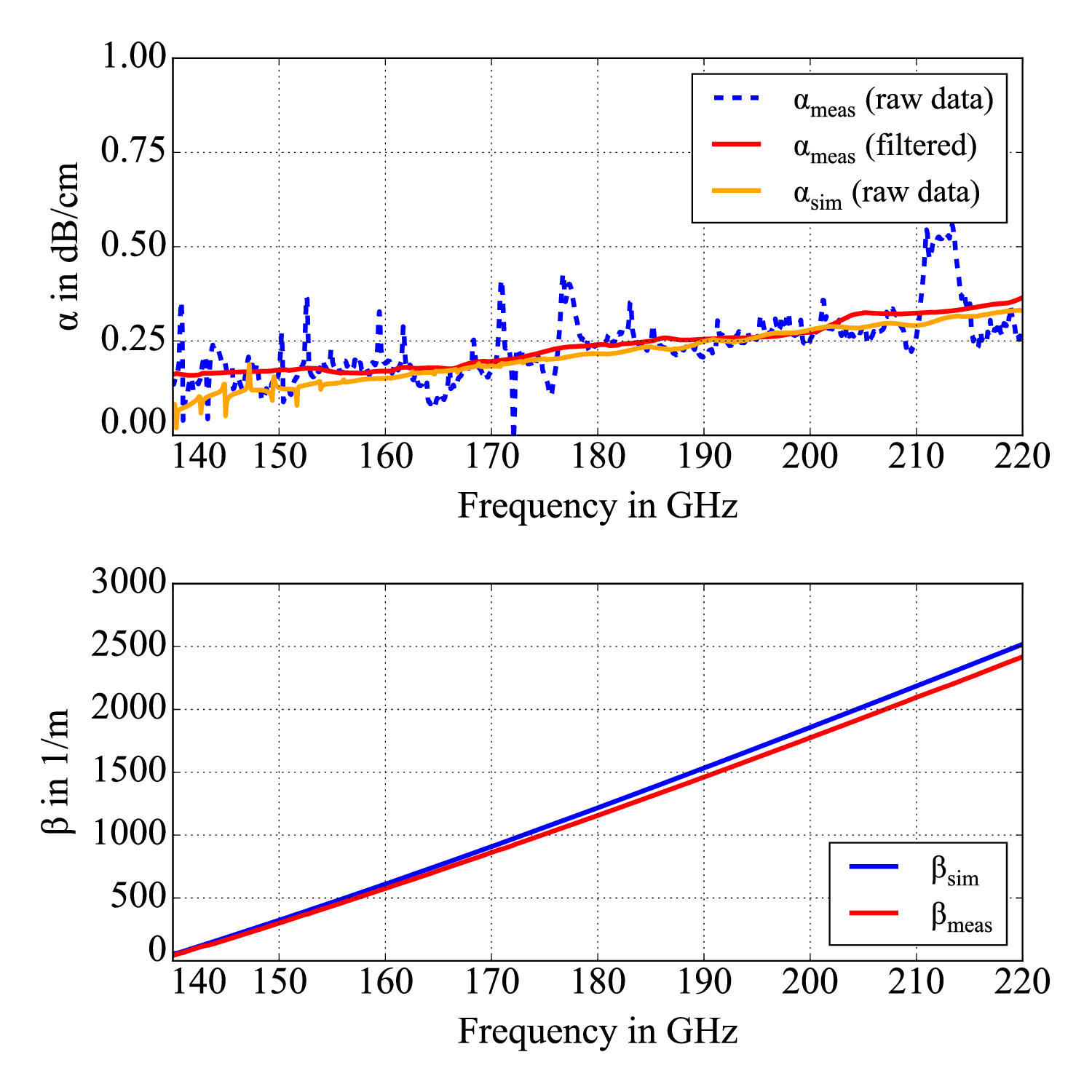}
\caption{Determination of the DIL's and propagation parameters $\alpha$ and $\beta$ with extraction of the mode-converters's influence on the transmission behavior.}
\label{fig:Gamma extraction of DILs}
\end{figure}

For the subsequent estimation of the attenuation- ($\alpha$) and phase coefficient ($\beta$), see Fig.~\ref{fig:Gamma extraction of DILs}, only the IL results of 2 different lengths are required and used. Herefore, a more thorough de-embedding technique based on all S-Parameter's eigenvalues is performed, enabling possible matching effects to be also taken into account and to be compensated for. The 3D-printed DILs show excellent consistency with a very low average attenuation of only \SI{0.25}{\frac{\decibel}{\centi\meter}}, making them suitable for low-loss transmission applications in the subTHz-domain. Due to the dielectric loss of the DIL and both ohmic loss and roughness of the copper surface, $\alpha$ is consistently rising, to values around \SI{0.35}{\frac{\decibel}{\centi\meter}} at \SI{220}{\giga\hertz}. The measured results and simulated expectations match in excellent accuracy, with only minor deviations in the \SI{140}{\giga\hertz} range. The phase coefficient $\beta$ shows an absolutely linear characteristic, which excludes phase distortions. The small deviation from the simulated $\beta$-value is caused by false assumption of the true effective permittivity value $\varepsilon_{\text{r,meas}} < \varepsilon_{\text{r,sim}}$. 
After determination of $\alpha_{\text{DIL}}$, the corresponding attenuation factor of the mode-converter is assessable. For an excitation of \SI{200}{\giga\hertz}, $\alpha_{\text{mc}}$ of one single mode-converter is only approx. \SI{0.3}{\decibel}, therefore the mode-converter can also be considered low-loss and has only minor influence on the measurement results. 
\begin{figure}[t]
\centering
\includegraphics[width=80mm]{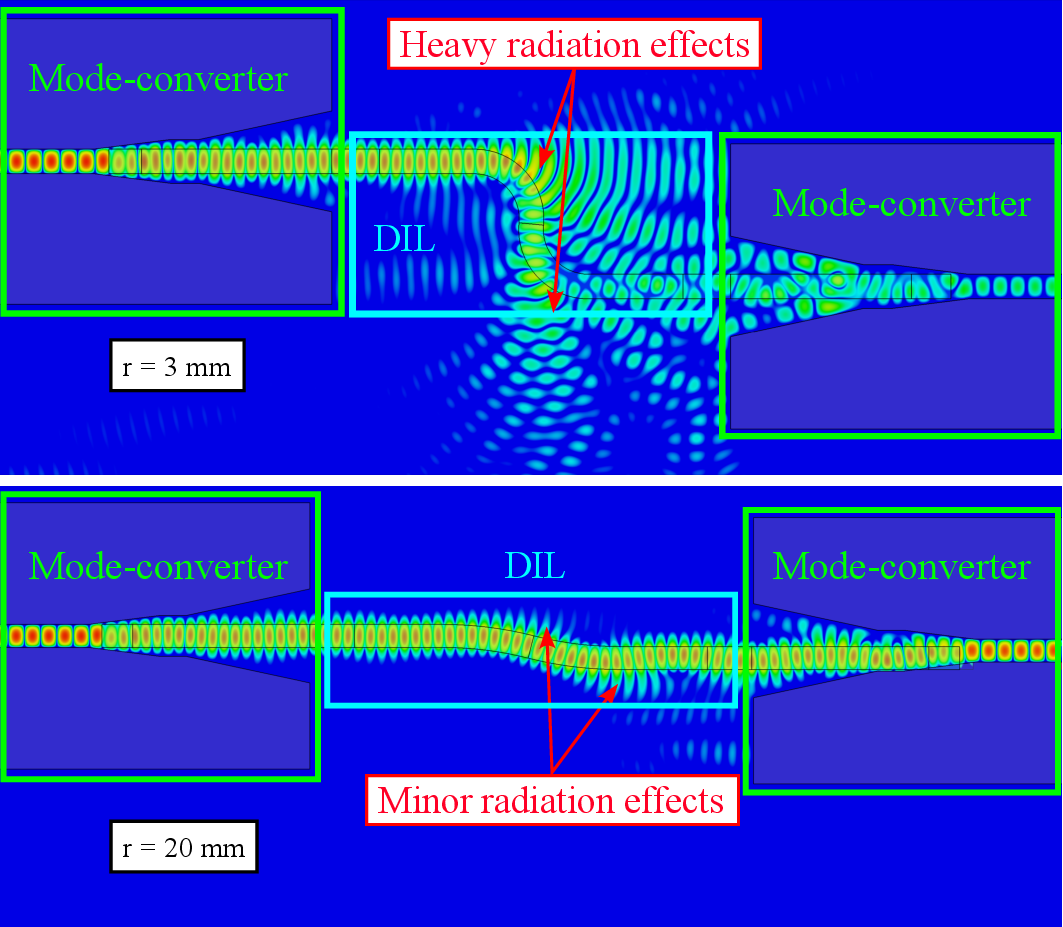}
\caption{CST simulation of the field distribution  for a  \SI{200}{\giga\hertz} excitation of DILs with narrow ($r = \SI{3}{\milli\meter}$, top) and wide ($r = \SI{20}{\milli\meter}$, bottom) bending radii, leading to variably strong parasitic radiation.}
\label{fig:DIL_BendingRadii_SIM}
\end{figure}
\begin{figure}[b]
\centering
\includegraphics[width=85mm]{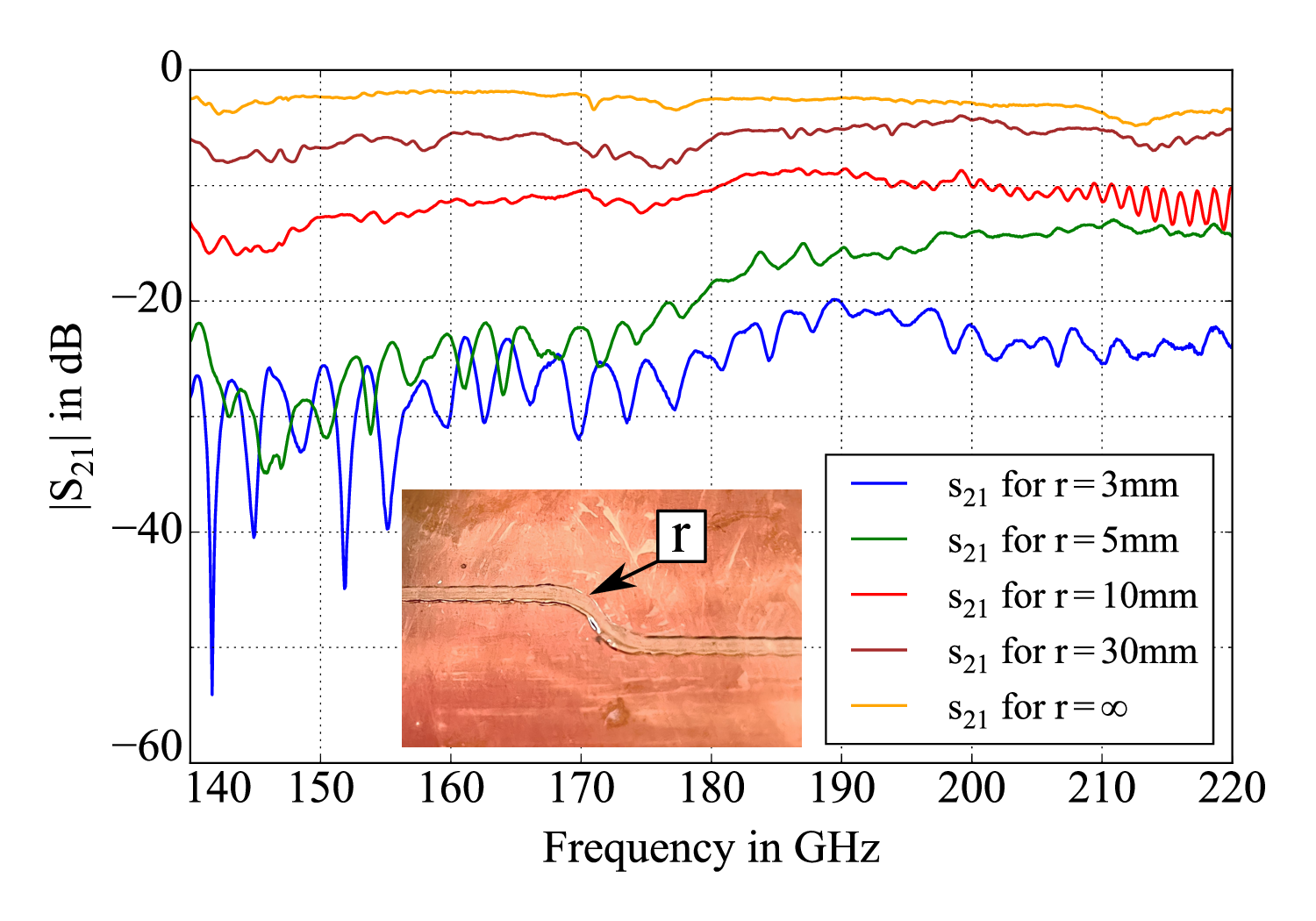}
\caption{Radii depedent measured insertion loss due to parasitic radiation effects at line discontinuities of DILs.}
\label{fig:DIL_BendingRadii_MEAS}
\end{figure}

Finally, since the characterized DILs will form the basis for complex dielectric networks and are to be used for subTHz interconnets that are not exclusively running straight, the influence of line discontinuities such as bending radii has to be investigated. As described in earlier fundamentals \cite{Distler.2018}, basically all types of discontinuities represent a sudden impedance variation for a dielectric waveguide, which interrupts the tight field binding and results in radiation at the location of the discontinuity, thus an increasing insertion loss can be expected. Fig.~\ref{fig:DIL_BendingRadii_SIM} shows the simulated field distribution for a  \SI{200}{\giga\hertz} excitation of two DILs with differing bending radii. For a narrow radius $r = \SI{3}{\milli\meter}$, the abrupt change of direction leads to heavy parasitic radiation at both corners. For a more gradual radius $r = \SI{20}{\milli\meter}$, only minor radiation effects are detected, therefore making this value much more suitable for line bends. For verification of this behavior, four DILs with differing bending radii are additively manufactured and characterized with the previously presented measurement setup, see Fig.~\ref{fig:DIL_BendingRadii_MEAS}. From the measured $s_{21}$, the radius-dependent insertion loss is clearly evident, which increases substantially for low values and prevents the use of DILs with bending radii $r < \SI{10}{\milli\meter}$. Radii $r > \SI{30}{\milli\meter}$ show only small deviations from straight lines ($r = \infty$), which means that they cause only a minor increase in the insertion loss and are therefore suitable for the desired low-loss applications. Furthermore, especially for small radii, a strong high-pass response can be observed, which is explained by the fact that the electromagnetic field is only loosely bound to its DIL at low frequencies due to the lower effective permittivity $\varepsilon_{\text{r,eff}}$, which particularly enhances the radiation effects and therefore the attenuation of those lower frequencies.

\section{Conclusion}
In this paper, dielectric image lines were additively manufactured by using exclusively 3D-printing FDM technology. A characterization of the transmission properties within a frequency converter setup proves the low-loss characteristic of the manufactured DILs with an average and maximum attenuation value of \SI{0.25}{\frac{\decibel}{\centi\meter}} and \SI{0.35}{\frac{\decibel}{\centi\meter}}, respectively. Moreover, the dielectric transmission line exhibits broadband matching in the entire frequency band from \SI{140}{\giga\hertz} to \SI{220}{\giga\hertz} with a return loss of \SI{20}{\decibel}, enabling the use of 3D-printed dielectric image lines for subTHz networks, especially for multi-channel signal distribution in chip-to-chip interconnects or for dielectric on-chip antennas. Feasibility of line bends is given, however, bending radii below \SI{30}{\milli\meter} are to be avoided due to the generation of strong parasitic radiation, that could lead to interference in adjacent components. With reasonable radii values, a compromise between additional insertion loss and flexibility and accessibility of the DIL routing could be reached. Unavoidable deviations of the desired DIL geometry have proven to have only minor impact on the quality of its transmission behavior. On the contrary, multiple tests have demonstrated that especially the initial taper, the accuracy of its mode-converter assembly and finally the surface quality of the underlying image plane are the dominant influence on the achievable transmission quality.

\section{Acknowledgement}
This work was supported by the German Research Foundation (DFG) within the frame of the research project TeraCaT (funding code 1519-54792 7558172).
The authors would also like to thank particularly the institutes' mechanics department, namely Kurt~Hack and Jürgen~Popp, for the precise fabrication of all necessary tools and assemblies.

\bibliography{Bibfile}{}
\bibliographystyle{IEEEtran}

\end{document}

%% file: IMS_modify_IEEEtran_18b_CTAN_V6.tex
\makeatletter

\def\@IMSauthorblockNAMEstyle{\normalfont\IMSauthorsize}
\def\@IMSauthorblockAFFILstyle{\normalfont\IMSaffilsize}
\def\@IMSauthorblockEMAILstyle{\normalfont\IMSaffilsize}
\def\IMSauthorblockNAME#1{%
\relax\@IMSauthorblockNAMEstyle%
#1%
}%
\def\IMSauthorblockAFFIL#1{%
\relax\@IMSauthorblockAFFILstyle%
\vskip\@IEEEauthorblockAtopspace
#1%
}%
\def\IMSauthorblockEMAIL#1{%
\relax\@IMSauthorblockEMAILstyle%
\vskip\@IEEEauthorblockAtopspace
#1%
}%
\newcommand{\IMSauthor}[1]{%
\ifIsBlindReviewVersion%
\author{\phantom{\parbox{\textwidth}{\center\relax#1}}}%
\else%
\author{\parbox{\textwidth}{\center\relax#1}}%
\fi%
}%
\newif\ifIsBlindReviewVersion

\def\IMSthispaperforfinalpublication{\IsBlindReviewVersionfalse}
\def\@maketitle{\newpage
\bgroup\par\addvspace{0.5\baselineskip}\centering%
\ifCLASSOPTIONtechnote% technotes
   {\bfseries\large\@IEEEcompsoconly{\sffamily}\@title\par}\vskip 1.3em{\lineskip .5em\@IEEEcompsoconly{\sffamily}\@author
   \@IEEEspecialpapernotice\par{\@IEEEcompsoconly{\vskip 1.5em\relax
   \@IEEEtitleabstractindextextbox{\@IEEEtitleabstractindextext}\par
   \hfill\@IEEEcompsocdiamondline\hfill\hbox{}\par}}}\relax
\else% not a technote
   \vskip0.2em{\IMStitlesize\ifCLASSOPTIONtransmag\bfseries\LARGE\fi\@IEEEcompsoconly{\sffamily}\@IEEEcompsocconfonly{\normalfont\normalsize\vskip 2\@IEEEnormalsizeunitybaselineskip
   \bfseries\Large}\@title\par}\vskip1.0em\par% CAUSAL PRODUCTIONS change on this line
   \ifCLASSOPTIONconference%
      {\@IEEEspecialpapernotice\mbox{}\vskip\@IEEEauthorblockconfadjspace%
       \mbox{}\hfill\begin{@IEEEauthorhalign}\@author\end{@IEEEauthorhalign}\hfill\mbox{}\par}\relax
   \else% peerreviewca, peerreview or journal
      \ifCLASSOPTIONpeerreviewca
         {\@IEEEcompsoconly{\sffamily}\@IEEEspecialpapernotice\mbox{}\vskip\@IEEEauthorblockconfadjspace%
          \mbox{}\hfill\begin{@IEEEauthorhalign}\@author\end{@IEEEauthorhalign}\hfill\mbox{}\par
          {\@IEEEcompsoconly{\vskip 1.5em\relax
           \@IEEEtitleabstractindextextbox{\@IEEEtitleabstractindextext}\par\hfill
           \@IEEEcompsocdiamondline\hfill\hbox{}\par}}}\relax
      \else% journal, peerreview or transmag
         \ifCLASSOPTIONtransmag
           {\@IEEEspecialpapernotice\mbox{}\vskip\@IEEEauthorblockconfadjspace%
            \mbox{}\hfill\begin{@IEEEauthorhalign}\@author\end{@IEEEauthorhalign}\hfill\mbox{}\par
           {\vspace{0.5\baselineskip}\relax\@IEEEtitleabstractindextextbox{\@IEEEtitleabstractindextext}\vspace{-1\baselineskip}\par}}\relax
         \else% journal or peerreview
           {\lineskip.5em\@IEEEcompsoconly{\sffamily}\sublargesize\@author\@IEEEspecialpapernotice\par
           {\@IEEEcompsoconly{\vskip 1.5em\relax
            \@IEEEtitleabstractindextextbox{\@IEEEtitleabstractindextext}\par\hfill
            \@IEEEcompsocdiamondline\hfill\hbox{}\par}}}\relax
         \fi
      \fi
   \fi
\fi\par\addvspace{0.0\baselineskip}\egroup}% CAUSAL PRODUCTIONS change on this line, reduce the vspace from 0.5\baselineskip to 0.0

\def\IMStitlesize{\@setfontsize{\IMStitlesize}{18}{21pt}}% CAUSAL PRODUCTIONS change on this line
\def\IMSauthorsize{\@setfontsize{\IMSauthorsize}{12}{13pt}}% CAUSAL PRODUCTIONS change on this line
\def\IMSaffilsize{\@setfontsize{\IMSaffilsize}{12}{13pt}}% CAUSAL PRODUCTIONS change on this line
\def\IMScaptionsize{\@setfontsize{\IMScaptionsize}{8}{9pt}}% CAUSAL PRODUCTIONS change on this line
\def\IMSbibsize{\@setfontsize{\IMSbibsize}{8}{9pt}}% CAUSAL PRODUCTIONS change on this line

\def\@IEEEauthorblockNstyle{\IMSauthorsize\@IEEEcompsocnotconfonly{\sffamily}\@IEEEcompsocconfonly{\large}}%CAUSAL PRODUCTIONS removed sublargesize to get correct IMSauthorsize
\def\@IEEEauthorblockAstyle{\IMSaffilsize\@IEEEcompsocnotconfonly{\sffamily}\@IEEEcompsocconfonly{\itshape}\@IEEEcompsocconfonly{\large}}%CAUSAL PRODUCTIONS removed normalsize to get correct IMSaffilsize
\def\@IEEEauthordefaulttextstyle{\IMSauthorsize\@IEEEcompsocnotconfonly{\sffamily}\sublargesize}%CAUSAL PRODUCTIONS

\def\thebibliography#1{\section*{\refname}%
    \addcontentsline{toc}{section}{\refname}%
    \IMSbibsize\@IEEEcompsocconfonly{\small}\vskip 0.3\baselineskip plus 0.1\baselineskip minus 0.1\baselineskip% CAUSAL PRODUCTIONS change on this line
    \list{\@biblabel{\@arabic\c@enumiv}}%
    {\settowidth\labelwidth{\@biblabel{#1}}%
    \leftmargin\labelwidth
    \advance\leftmargin\labelsep\relax
    \itemsep \IEEEbibitemsep\relax
    \usecounter{enumiv}%
    \let\p@enumiv\@empty
    \renewcommand\theenumiv{\@arabic\c@enumiv}}%
    \let\@IEEElatexbibitem\bibitem%
    \def\bibitem{\@IEEEbibitemprefix\@IEEElatexbibitem}%
\def\newblock{\hskip .11em plus .33em minus .07em}%
\ifCLASSOPTIONtechnote\sloppy\clubpenalty4000\widowpenalty4000\interlinepenalty100%
\else\sloppy\clubpenalty4000\widowpenalty4000\interlinepenalty500\fi%
    \sfcode`\.=1000\relax}

\long\def\@makecaption#1#2{%
\ifx\@captype\@IEEEtablestring%
\par\@IEEEtabletopskipstrut% strut used to align table caption with facing column
\else
\@IEEEfigurecaptionsepspace
\fi
\setbox\@tempboxa\hbox{\normalfont\IMScaptionsize {#1.}\nobreakspace\nobreakspace #2}%
\ifdim \wd\@tempboxa >\hsize%
\setbox\@tempboxa\hbox{\normalfont\IMScaptionsize {#1.}\nobreakspace\nobreakspace}%
\parbox[t]{\hsize}{\normalfont\IMScaptionsize\noindent\unhbox\@tempboxa#2}%
\else
\ifCLASSOPTIONconference \hbox to\hsize{\normalfont\IMScaptionsize\hfil\box\@tempboxa\hfil}%
\else \hbox to\hsize{\normalfont\IMScaptionsize\box\@tempboxa\hfil}%
\fi\fi
\ifx\@captype\@IEEEtablestring%
\@IEEEtablecaptionsepspace
\else
\fi}

\newlength\tablecaptiontotableskip
\newlength\figuretocaptionskip
\def\@IEEEfigurecaptionsepspace{\vskip\figuretocaptionskip\relax}%
\def\@IEEEtablecaptionsepspace{\vskip\tablecaptiontotableskip\relax}%

\def\abstract{\normalfont%
\@IEEEabskeysecsize\bfseries\textit{\abstractname}\,\bfseries\textit{---}\,%
\@IEEEgobbleleadPARNLSP}%

\def\IEEEkeywords{\normalfont%
\@IEEEabskeysecsize\bfseries\textit{\IEEEkeywordsname}\,\bfseries\textit{---}\,%
\@IEEEgobbleleadPARNLSP}%
\def\endIEEEkeywords{\relax\vspace{0.67ex}%
\par\if@twocolumn\else\endquotation\fi%
\normalsize\normalfont}%

\DeclareRobustCommand*{\IMSauthorrefmark}[1]{\raisebox{0pt}[0pt][0pt]{\textsuperscript{\footnotesize{#1}}}}%
\def\@IEEEauthorblockNtopspace{0ex}
\def\@IEEEauthorblockAtopspace{1mm}
\def\tablename{Table}% IMS: was TABLE in IEEETran, but we are now using capitalized caption text not smallcaps
\def\thetable{\arabic{table}}% IMS: Tables are numbered using arabic not roman
\def\IEEEkeywordsname{Keywords}% use Keywords instead of Index Terms
\def\subsubsection{\@startsection{subsubsection}{3}{\z@}{1.5ex plus 1.5ex minus 0.5ex}%
{0.7ex plus .5ex minus 0ex}{\normalfont\normalsize\itshape}}%
\def\@seccntformat#1{\csname the#1dis\endcsname\relax}% moved the spacer \hskip 0.5em to individual handlers below
\def\thesectiondis{\thesection.\hskip 0.5em}%					I.	% CAUSAL PRODUCTIONS: moved the hskip spacer from \@seccntformat to here
\def\thesubsectiondis{{\hbox to\parindent{\Alph{subsection}.}}}%		B.	% CAUSAL PRODUCTIONS: indent the subsection name to match paragraph indent
\def\thesubsubsectiondis{{\hbox to \parindent{\arabic{subsubsection})}}}%	3)	% CAUSAL PRODUCTIONS: indent the subsubsection name to match paragraph indent
\def\theparagraphdis{{\hbox to \parindent{\alph{paragraph})}}}%			d)	% CAUSAL PRODUCTIONS: indent the subsubsubsection name to match paragraph indent
\IEEEilabelindentA \parindent
\IEEEilabelindent \IEEEilabelindentA
\IEEEelabelindent \parindent
\IEEEdlabelindent \parindent
\IEEElabelindent \parindent
\newlength\@IMSparindent
\newcommand\IMSdisplayacksection[1]{%
\ifIsBlindReviewVersion%
\noindent\phantom{\parbox[t]{\columnwidth}{\normalbaselines\setlength{\parindent}{\@IMSparindent}{#1}\strut}}%\IMSacktext
\else%
\noindent\parbox[t]{\columnwidth}{\normalbaselines\setlength{\parindent}{\@IMSparindent}{#1}\strut}%
\fi%
}%

\makeatother